# Calculation of the Electron Swarm Parameters in He/H$_2$ and Ne/H$_2$ Mixtures

Mohamed Mostefaoui*, and Djilali Benyoucef
Laboratoire Génie Electrique et Energies Renouvelables, Université Hassiba Benbouali de Chlef,
Chlef, Algeria
*ostefm@gmail.com

*Abstract*— The knowledge of all the electron swarm parameters as a function of the reduced electric field is indispensable for the fluid modeling of electrical discharge at low pressure. In this paper we present the evolution of the swarm parameters as a function of the reduced electric field in the He/H$_2$ and Ne/H$_2$ mixtures. For this we use a BE solver specially designed for this purpose, freely available under the name BOLSIG+, which is more general and easier to use than most other BE available solvers. The solver provides a stable solution state of the BE for the electrons in a uniform electric field, it is based on the conventional two-term approximation. The obtained results are validated through a comparison to experimental swarm parameters already exist in the literature.

*Keywords*— *cross sections; mobility; diffusion; ionisation coefficient; swarm parameters; Boltzmann equation*

## I. Introduction

Fluid models of plasma discharge describe the transport of electrons, ions and other reactive or excited species by the moments of Boltzmann equation (BE). Generally the mass continuity equation and the momentum continuity equation in the case of local electric field balance. In addition, the electron energy equation should also be included in the case of local energy balance. These equations are function of swarm parameters, which are: the mobility, the diffusion, and the ionization coefficient by electrons impact. The values of these parameters are the input data of the fluid models [1–4]. The electron swarm parameters are strongly related to the electron energy distribution function (EEDF); unfortunately, in plasma discharges, this function is not always Maxwellian, it usually depends on the pressure of the background gas. Some authors prefer to measure and tabulate the swarm parameters as a function of the reduced electric field $E/N$ (ratio of the electric field to the background gas density), then use them as input data in the fluid models [5]. The current approach is to solve the Boltzmann equation for a series of reduced electric field values and to take the resulting calculated swarm parameters as a function of the reduced electric field or the electrons average energy, which are then interpolated for easy used in fluid models. So, the electron swarm parameters can be calculated through the electron energy distribution function (EEDF), which can be obtained by solving the Boltzmann Equation (BE). However, to solve the Boltzmann equation requires knowledge of the basic data of all the elementary collision processes such as elastic and inelastic collisions (excitation, ionization, dissociation, etc.). These processes can be described by their cross-sections data [6], these last data can be used directly as input for the particle models [7-10]. In the literature, there are several works concerning the computation of electron swarm parameters from solving the Boltzmann equation [11–16]. The initial goal of these techniques of Boltzmann equation solving was the simulation of the experiment studies with high numerical precision. In order to obtain Boltzmann equation solving satisfies the fluid models conditions, the BE solver must have the following objectives:

1- Take into account a large range of discharge conditions as reduced electric field, ionization degree, gas temperature, etc.

2- The calculated swarm parameters (mobility, diffusion, and the ionization coefficient) should correspond to the same coefficients appearing in the first moments of the Boltzmann equation.

3- The computation errors of the swarm parameters must not influence the accuracy of the fluid model (less than 0.1%).

4- The solver should be fast and reliable and without adjustable parameters.

In order to solve the Boltzmann equation, there are several solvers which are often used by the authors in the field of the electric discharges modeling; in particular the commercial software ELENDIF [15] and the free software BOLSIG+ [17].

## II. Boltzmann Equation Solving and Swarm Parameters Calculation

### A. Boltzmann Equation

The electrons Boltzmann equation in an ionized gas is given by the following equation:

$$\frac{\partial f}{\partial t} = \mathbf{v}.\nabla f - \frac{e}{m_e}\mathbf{E}.\nabla_v f = C[f]$$

$$\nabla_v = \frac{\partial}{\partial v_x}\mathbf{i} + \frac{\partial}{\partial v_y}\mathbf{j} + \frac{\partial}{\partial v_z}\mathbf{k} \quad (1)$$

where $f$ is the electron distribution in six-dimensional phase space, $\mathbf{v}$ is the velocity vector, $e$ is the elementary charge, $m_e$ is the electron mass, $\mathbf{E}$ is electric field vector, $C[f]$ represents the rate of electrons production and loss due to the collisions, and $v_x$, $v_y$, $v_z$ are the velocity coordinates. To solve this last equation, we must make the following simplifications:

1- The collisions probabilities and the electric field are considered uniform in the ordinary space.

2- The electrons distribution function $f$ is considered symmetric in velocity space around the electric field direction.

3- In ordinary space $f$ can vary only in the direction of the electric field, which is in the $z$-axis direction.

By using the spherical coordinates in the velocity space, we can obtain the following equation:

$$\frac{\partial f}{\partial t} + \upsilon \cos\theta \frac{\partial f}{\partial z} - \frac{e}{m_e} E \left( \cos\theta \frac{\partial f}{\partial \upsilon} + \frac{\sin^2\theta}{\upsilon} \frac{\partial f}{\partial \cos\theta} \right) = C[f] \quad (2)$$

where $\upsilon$ is the velocity magnitude, $\theta$ is the angle between the velocity and the electric field direction and $z$ is the position along this direction. The electrons distribution function $f$ depend on four coordinates: $\upsilon$, $\theta$, $t$ et $z$. Now to simplify the dependence with $\theta$, we use the classical two-term approximation method, and to simplify the temporal dependence, we consider that the electric field and the electron distribution function are stationary. Additional exponential dependence of f on t or on z is assumed to account for electron production or loss due to ionization and attachment.

*B. Two-term approximation method*

A commonly approach to solving equation (2) is to develop $f$ in terms of Legendre polynomials of $\cos\theta$, and then to construct from equation (2) a system of equations for expansion coefficients. It is necessary to use more than six expansion terms to obtain results with very great precision [12]. BOLSIG and ELENDIF software use a two-term approximation and they give useful results [18]. For high values of reduced electric fields (E/N > 1000 Td), the electron distribution function becomes more and more anisotropic due to the increase in the number of inelastic collisions in comparison to the number of the elastic collisions [19], but the computational errors in the evaluation of the swarm parameters remain acceptable for fluid modeling. By using two terms approximation, the electron distribution function can be written as follow:

$$f(\upsilon, \cos\theta, z, t) = f_0(\upsilon, z, t) + f_1(\upsilon, z, t)\cos\theta \quad (3)$$

where $f_0$ is the isotropic part of $f$, and $f_1$ is the anisotropic part due to presence of electric field. Note that $\theta$ is defined with respect to the electric field direction, in this case $f_1$ is negative, some authors taken $f_1$ in the direction of the drift velocity, and in this last case $f_1$ is positive. When $f$ is stationary and uniform in the ordinary space (independent of $z$ and $t$), it can normalized as follow:

$$\iiint f \, d^3\upsilon = 4\pi \int_0^\infty f_0 \upsilon^2 d\upsilon = n(z,t) = n \quad (4)$$

where $n$ is the electron density number. By substituting the equation of f (eq.3) in equation (2) and integrating over $\cos\theta$, equation (2) becomes as follows:

$$\frac{\partial f_0}{\partial t} + \frac{\gamma}{3}\left[ \sqrt{\varepsilon} \frac{\partial f_1}{\partial z} - \frac{1}{\sqrt{\varepsilon}} \frac{\partial}{\partial \varepsilon}(\varepsilon E f_1) \right] = C_0$$

$$\frac{\partial f_1}{\partial t} + \gamma\sqrt{\varepsilon}\left[ \frac{\partial f_0}{\partial z} - E\frac{\partial f_0}{\partial \varepsilon} \right] = -N\sigma_m \sqrt{\varepsilon} f_1 \quad (5)$$

where $\gamma = (2e/m)^{1/2}$, $\varepsilon = (\upsilon/\gamma)^2$ is the electron energy in electronvolts, $C_0$ is the rate change in $f_0$ due to the collisions, $N$ is the gas density, and $N\sigma_m \varepsilon^{1/2} f_1$ is the total momentum-transfer cross-section $\sigma_m$ of all possible collision processes k with gas particles.

$$\sigma_m = \sum_k p_k \sigma_k \quad (6)$$

where $p_k$ is the molar fraction of the target particles, including the case where the background gas is a mixture, $\sigma_k$ is the momentum transfer cross section of process $k$ [20], for the inelastic collision $\sigma_k$ represents the total cross section.

*C. Electron density growth*

Due to the ionization and attachment collisions, the total number of electrons is not conserved, and consequently the electron distribution function $f$ cannot be uniform and stationary. [18, 20–22]. Following the technique proposed in references, we can separate the energy-dependence of f from its dependence on time and space by taking:

$$f_{0,1}(\varepsilon, z, t) = \frac{1}{2\pi\gamma^3} F_{0,1}(\varepsilon) n(z, t) \quad (7)$$

where the electron energy distribution function is independent of spatial and temporal coordinates, and normalized by:

$$\int_0^\infty \sqrt{\varepsilon} F_0(\varepsilon) d\varepsilon = 1 \quad (8)$$

Now, the electrons density depends on the rate of the electrons production, and by basing on the exponential temporal growth without spatial dependence, the temporal growth rate of the electron density is equal to the net generation frequency ($v_i$) given by the following equation:

$$\frac{1}{n_e}\frac{\partial n_e}{\partial t} = \overline{v_i} = N\gamma \int_0^\infty \left( \sum_{k_i} p_{k_i}\sigma_{k_i} - \sum_{k_a} p_{k_a}\sigma_{k_a} \right) \varepsilon F_0 d\varepsilon \quad (9)$$

where $k_i$ indicate ionization process and $k_a$ indicate attachment process. From equation (5), the anisotropic part of the electron energy distribution function becomes:

$$F_1 = \frac{E}{N}\frac{1}{\tilde{\sigma}_m}\frac{\partial F_0}{\partial \varepsilon}$$

$$\tilde{\sigma}_m = \sigma_m + \frac{\overline{v_i}}{N\gamma\sqrt{\varepsilon}} \quad (10)$$

By substituting $F_1$ (eq.10) in isotropic part of equation (5), we find:

$$-\frac{\gamma}{3}\frac{\partial}{\partial \varepsilon}\left[\left(\frac{E}{N}\right)^2 \frac{\varepsilon}{\tilde{\sigma}_m}\frac{\partial F_0}{\partial \varepsilon}\right] = \tilde{C}_0 + \tilde{R} \quad (11)$$

where the collision term is given by the following expression:

$$\tilde{C}_0 = 2\pi\gamma^3 \sqrt{\varepsilon}\,\frac{C_0}{Nn}$$
$$\tilde{R} = -\frac{\bar{v}_i}{N}\sqrt{\varepsilon}\,F_0 \quad (12)$$

This last term represents the energy required to heat the secondary electrons to the average electron energy.

Now, we rely on exponential spatial growth without temporal dependence, where the electrons move in the opposite direction of the electric field, and their flux and density increase exponentially with a constant spatial growth coefficient $\alpha$ (the first Townsend coefficient), the latter appears in the fluid model equations, and it is defined as follows:

$$\alpha = -\frac{1}{n}\frac{\partial n}{\partial z} = -\frac{\bar{v}_i}{\omega} \quad (13)$$

where the average velocity $\omega$ is determined by $F_1$, it is constant in space with a negative value. By substituting $\alpha$ (eq.13) in anisotropic part of equation (5), we find:

$$F_1 = -\frac{1}{\sigma_m}\left(\frac{E}{N}\frac{\partial F_0}{\partial \varepsilon} + \frac{\alpha}{N}F_0\right) \quad (14)$$

Finally, the isotropic part of the electron energy distribution function (eq.5) can be written as follows:

$$-\frac{\gamma}{3}\frac{\partial}{\partial \varepsilon}\left[\left(\frac{E}{N}\right)^2 \frac{\varepsilon}{\sigma_m}\frac{\partial F_0}{\partial \varepsilon}\right] = 2\pi\gamma^3\sqrt{\varepsilon}\,\frac{C_0}{Nn} + \frac{\alpha}{N}\frac{\gamma}{3}\left[\frac{\varepsilon}{\sigma_m}\left(2\frac{E}{N}\frac{\partial F_0}{\partial \varepsilon}\frac{\alpha}{N}F_0\right) + \frac{E}{N}F_0\frac{\partial}{\partial \varepsilon}\left(\frac{\varepsilon}{\sigma_m}\right)\right] \quad (15)$$

The first Townsend coefficient $\alpha$ can be found by combining equation (13) and equation (14):

$$\omega = -\frac{1}{3}\gamma\int_0^\infty F_1 \varepsilon \,d\varepsilon = -\mu E + \alpha D = -\frac{\bar{v}_i}{\alpha} \quad (16)$$

This leads to:

$$\alpha = \frac{1}{2D}\left(\mu E - \sqrt{(\mu E)^2 - 4D\bar{v}_i}\right) \quad (17)$$

where $\mu$ et $D$ are respectively, the mobility and the diffusion coefficient, then, the reduced mobility $\mu N$ and characteristic energy (reduced diffusion $DN$) can be calculated by the following expressions:

$$\mu N = -\frac{\gamma}{3}\int_0^\infty \frac{\varepsilon}{\tilde{\sigma}_m}\frac{\partial F_0}{\partial \varepsilon}d\varepsilon \quad (18)$$

$$DN = \frac{\gamma}{3}\int_0^\infty \frac{\varepsilon}{\tilde{\sigma}_m}F_0\,d\varepsilon \quad (19)$$

## III. RESULTS AND DISCUSSION

The cross sections of Helium, Neon, and Hydrogen used in this work are those of Phelps [23], the results are obtained for a stationary electric field. The calculated values are compared with the measurements already exists in the literature.

### A. He/H$_2$ Mixture

The following figure shows the mobility, the characteristic energy, and the ionization coefficient in the He/H$_2$ mixture as a function of the reduced electric field $E/N$ under the following conditions:

1: He 100%/H$_2$ 0%
2: He 75%/H$_2$ 25%
3: He 50%/H$_2$ 50%
4: He 25%/H$_2$ 75%
5: He 0%/H$_2$ 100%.

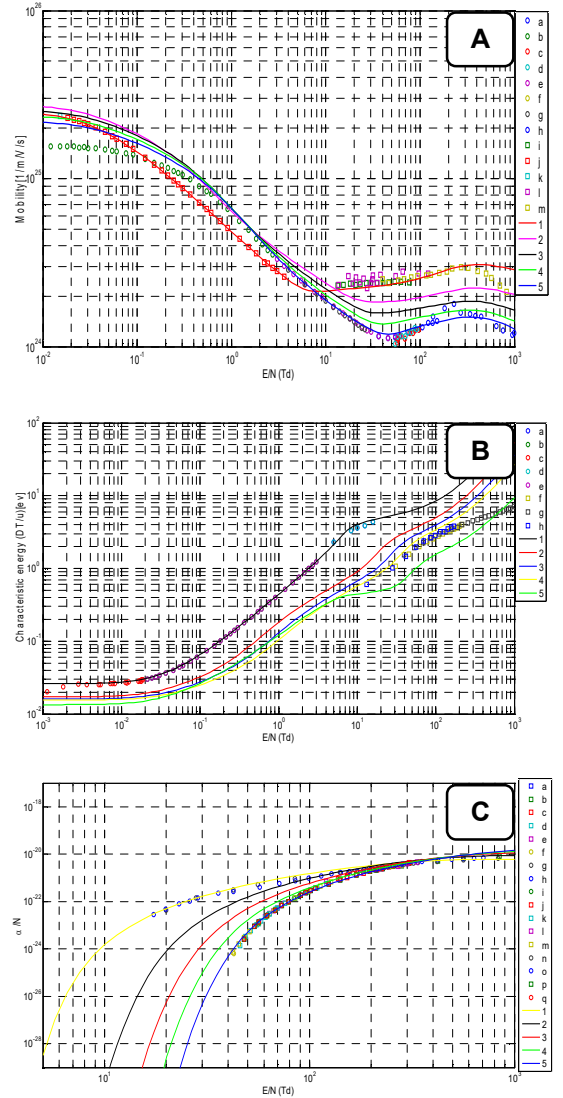

Fig.1. Comparison between the calculated swarm parameters and those measured in the mixture (He/H$_2$); (A): Electron mobility, (B): The characteristic energy of electrons, (C) Ionization coefficient

*References of the measurements data in figure* 1.A

  a: Pak and Phelps (1961) [24]

  b: Crompton *et al.* (1967) [25]

  c: Anderson (1964) [26]

  d: Phelps *et al.* (1960) [27]

  e: Stern (1963) [28]

  f: Lowke (1963) [29]

  g: Pak and Phelps (1961) [24]

  h: Robertson (1971) [30]

  i: Rose (1956); [31]

  j: Schlumbohm (1965) [32]

  k: Virr *et al.* (1972) [33]

  l: Wagner *et al.* (1967) [34]

  m: Warren and Parker (1962) [35]

*References of the measurements data in figure* 1.B

  a : Warren and Parker (1962) [35]

  b : Crompton *et al.* (1967) [25]

  c: Al-Amin and Lucas (1967) [36]

  d : Towsend and Bailey (1923) [37]

  e : Towsend and Bailey (1923) [38]

  f: Crompton *et al.* 1968 [39]

  g: Lawson, J. Lucas 1965 [40]

  h: k: Virr *et al.* (1972) [33]

*References of the measurements data in figure* 1.C

  a: Chanin and Rork (1964) [41]

  b: Davies *et al.* (1962) [42]

  c: Dunlop (1949) [43]

  d: Abdelnabi and Massey (1953) [44]

  e: Lakshminarasimha *et al.* (1975) [45]

  f: Townsend and MacCallum (1934) [46]

  g: Barna *et al.* (1964) [47]

  h: Blasberg and de Hoog (1971) [48]

  i: Blevin *et al.* 1957[49]

  j: Crompton *et al.* 1956 [50]

  k: Chanin and Rork (1963) [51]

  l: Rose *et al.* 1956[52]

  m: Cowling J. Fletcher (1973) [53]

  n: Frommhold (1960) [54]

  p: Folkard and Haydon (1971) [55]

  q: Haydon and Robertson (1961) [56]

  r: Hopwood *et al.* 1956[57]

### B. Ne/H$_2$ Mixture

In Figure 2, we show the mobility, the characteristic energy, and the ionization coefficient in the Ne/H$_2$ mixture as a function of the reduced electric field *E*/*N* under the following conditions:

1: Ne100%/H$_2$ 0%
2: Ne 75%/H$_2$ 25%
3: Ne 50%/H$_2$ 50%
4: Ne 25%/H$_2$ 75%
5: Ne 0%/H$_2$ 100%

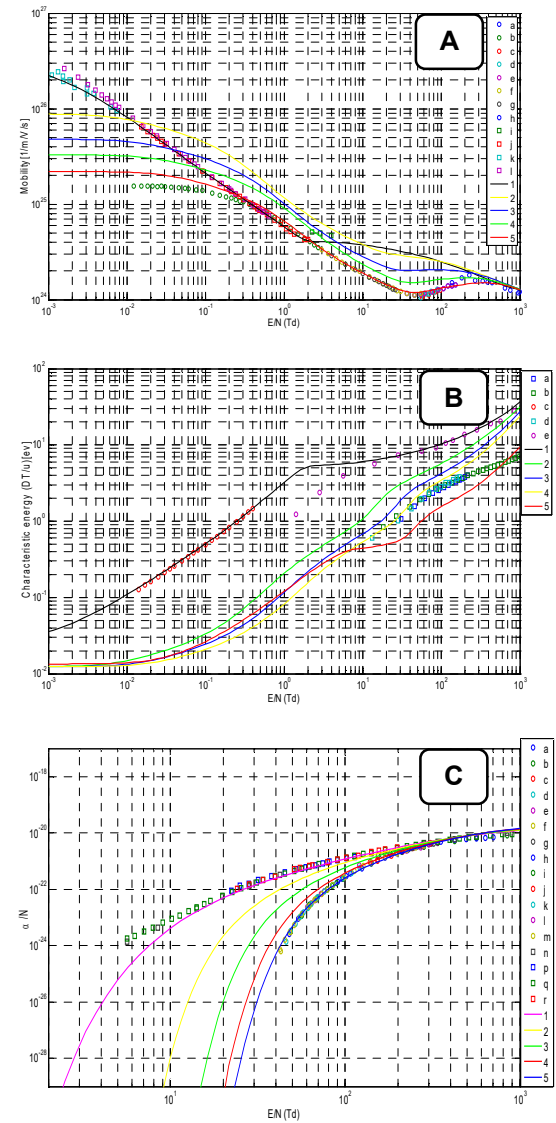

Fig.2. Comparison between the calculated swarm parameters and those measured in the mixture (Ne/H$_2$); (A): Electron mobility, (B): The characteristic energy of electrons, (C) Ionization coefficient

*References of the measurements data in figure* 2.A

    a: Pak and Phelps (1961) [24]

    b: Roberston (1972) [58]

    c: Nielsen (1936) [59]

    d: Bowe (1960) [60]

    e: Lowke (1963) [29]

    f: Pak and Phelps (1961) [24]

    g: Robertson (1971) [30]

    h: Rose (1956); [31]

    i: Schlumbohm (1965) [32]

    j: Virr *et al.* (1972) [33]

    k: Wagner *et al.* (1967) [34]

    l: Warren and Parker (1962) [35]

*References of the measurements data in figure* 2.B

    a: Al-Amin and Lucas (1967) [36]

    b: Koizumi *et al.* (1984) [61]

    c: Crompton *et al.* 1968 [39]

    d: Lawson, J. Lucas 1965 [40]

    e: k: Virr *et al.* (1972) [33]

*References of the measurements data in figure* 2.C

    a : k: Chanin and Rork (1963) [51]

    b : Dutton *et al.* (1969) [62]

    c : Kruithof and Druyvesteyn (1937) [63]

    d : Willis and Morgan (1968) [64]

    e: Kruithof (1940) [65]

    f: Bhattacharya (1976) [66]

    g: Barna *et al.* (1964) [47]

    h: Blasberg and de Hoog (1971) [48]

    i: Blevin *et al.* 1957 [49]

    j: Crompton *et al.* 1956 [50]

    k: Chanin and Rork (1963) [51]

    l: Rose *et al.* 1956 [52]

    m: Cowling J. Fletcher (1973) [53]

    n: Frommhold (1960) [54]

    p: Folkard and Haydon (1971) [55]

    q: Haydon and Robertson (1961) [56]

    r: Hopwood *et al.* 1956 [57]

The results obtained for the pure gases of Helium, Neon, and Hydrogen are in very good agreement with the measurements data, where for low electric field the electron mobility, the characteristic energy, and also the ionization coefficient decrease by increasing the ratio of hydrogen, this is can be explained by the lower excited state of hydrogen (rotational states and vibrational states). These last, absorbs the electrical energy gained by the electrons movement in the electric field, this decrease is due as a result of inelastic collisions with the hydrogen molecule.

IV. CONCLUSION

In this work we have shown the evolution of the swarm parameters in He/$H_2$ and Ne/$H_2$ mixtures. The results obtained for the pure gases of Helium, Neon, and Hydrogen are in very good agreement with the measurements. In general, we note that the ionization coefficient increases with the increase of the percentage of the noble gases (He, Ne). For this reason, these gases are used as additive with the any gas to increase the density of the electrons in the electric discharges.